\documentclass{article}
\usepackage[preprint]{neurips_2025}
\usepackage[utf8]{inputenc} 
\usepackage[T1]{fontenc}    
\usepackage{hyperref}       
\usepackage{url}            
\usepackage{booktabs}       
\usepackage{amsfonts}       
\usepackage{nicefrac}       
\usepackage{microtype}      

\usepackage{caption}
\usepackage[table,xcdraw]{xcolor}
\usepackage{graphicx}
\usepackage{array}
\usepackage[utf8]{inputenc}
\usepackage{booktabs} 
\usepackage{multirow} 
\usepackage{float}
\usepackage{makecell}
\usepackage{amsmath}
\usepackage{algorithmic}
\usepackage{algorithm}
\usepackage{tcolorbox}
\usepackage{tikz}
\usepackage{fontawesome5}
\usepackage{enumitem}
\setlist[itemize]{itemsep=0.5pt, topsep=2pt}

\title{Kling-Foley: Multimodal Diffusion Transformer for High-Quality Video-to-Audio Generation}

\author{
\textbf{Jun Wang}*, \textbf{Xijuan Zeng}*, \textbf{Chunyu Qiang}*, 
\textbf{Ruilong Chen}, \textbf{Shiyao Wang}, \textbf{Le Wang}, \\
\textbf{Wangjing Zhou}, \textbf{Pengfei Cai}, \textbf{Jiahui Zhao}, 
\textbf{Nan Li}, \textbf{Zihan Li}, \textbf{Yuzhe Liang}, \\
\textbf{Xiaopeng Wang}, \textbf{Haorui Zheng}, \textbf{Ming Wen}, 
\textbf{Kang Yin}, \textbf{Yiran Wang}, \\
\textbf{Nan Li}, \textbf{Feng Deng}, \textbf{Liang Dong}, 
\textbf{Chen Zhang}, \textbf{Di Zhang}, \textbf{Kun Gai} \\[1.2ex]
Kuaishou Technology, Beijing, China \\[1.2ex]
\footnotesize\texttt{\{wangjun06,zengxijuan,qiangchunyu\}@kuaishou.com}
}

\begin{document}

\maketitle

\begin{abstract}
  We propose Kling-Foley, a large-scale multimodal Video-to-Audio generation model that synthesizes high-quality audio synchronized with video content. In Kling-Foley, we introduce multimodal diffusion transformers to model the interactions between video, audio, and text modalities, and combine it with a visual semantic representation module and an audio-visual synchronization module to enhance alignment capabilities. Specifically, these modules align video conditions with latent audio elements at the frame level, thereby improving semantic alignment and audio-visual synchronization. Together with text conditions, this integrated approach enables precise generation of video-matching sound effects. In addition, we propose a universal latent audio codec that can achieve high-quality modeling in various scenarios such as sound effects, speech, singing, and music.  We employ a stereo rendering method that imbues synthesized audio with a spatial presence. At the same time, in order to make up for the incomplete types and annotations of the open-source benchmark, we also open-source an industrial-level benchmark Kling-Audio-Eval. Our experiments show that Kling-Foley trained with the flow matching objective achieves new audio-visual SOTA performance among public models in terms of distribution matching, semantic alignment, temporal alignment and audio quality. Homepage is available at: \url{https://klingfoley.github.io/Kling-Foley/}
\end{abstract}

\renewcommand{\thefootnote}{\fnsymbol{footnote}} 
\footnotetext[1]{Equal contributions.}

\begin{figure*}[h]
 \centering
 \includegraphics[width=\linewidth]{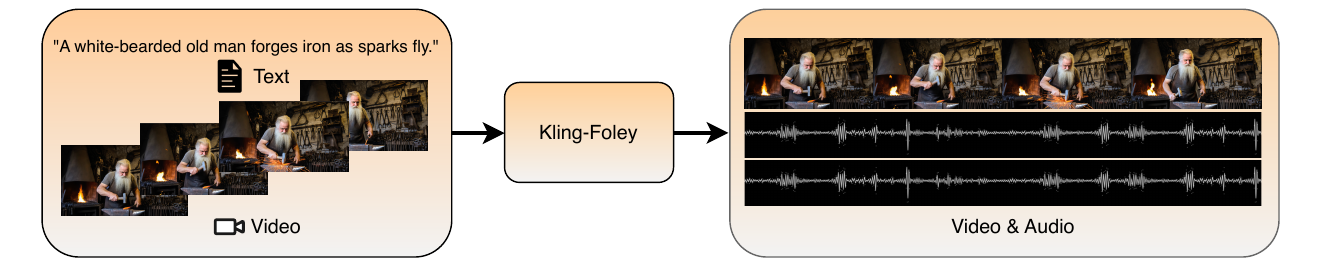}
 \caption{We propose Kling-Foley, a large-scale multimodal Video-to-Audio generation
model. Taking an input video and an optional text prompt, the model synthesizes high-fidelity audio that is semantically aligned and temporally synchronized with the video content, encompassing elements such as sound effects and background music. Significantly, Kling-Foley can produce audio sequences of arbitrary duration, dynamically adapting to the length of the input video. }
 \label{fig:abstract}
\end{figure*}

\section{Introduction}

Video generation has emerged as a focal point in generative AI research, with many models producing visually convincing results \cite{ kong2024hunyuanvideo,polyak2024movie,kuaishou2024kling,videoworldsimulators2024}. However, these generated videos remain silent, requiring separate post-production audio dubbing to align with human perception and real-world expectations. Manual dubbing proves costly, inefficient, and heavily reliant on specialized expertise.  

To automate this process, Text-to-Audio (TTA) models generate non-speech audio \cite{Kreuk2022AudioGenTG, Yang2022DiffsoundDD, Liu2023AudioLDM2L, Ghosal2023TexttoAudioGU, Huang2023MakeAnAudio2T}, such as  sound effects and background music from text descriptions. The core objective of TTA involves translating natural language into high-fidelity, semantically aligned audio signals. However, TTA faces critical limitations: since it processes text alone, the generated audio often not temporally or semantically aligned with the video content. For instance, the footsteps may not match the shoe movements (temporal misalignment), or a "rain" description might yield light rain sounds while the video shows a downpour (semantic misalignment).   

To resolve these issues, Video-to-Audio (V2A) models leverage both text descriptions and visual input\cite{Xing2024SeeingAH, Wang2023V2AMapperAL, Xie2024SonicVP, Jeong2024ReadWA, Zhou2017VisualTS}. Unlike TTA frameworks, V2A systems typically extract frame-level video features via pre-trained encoders to supplement textual inputs. While many V2A approaches extend pre-trained TTA models, their performance suffers due to the scarcity of high-quality, triple-modality (video-audio-text) datasets. Consequently, existing models often exhibit poor temporal/semantic alignment and low-fidelity audio output. MM-Audio\cite{mmaudio}, the current state-of-the-art (SOTA) model, addresses these weaknesses via a multimodal joint-training paradigm. Nevertheless, its reliance on limited open-source datasets restricts it to ambient sound and sound effects, excluding capabilities like background music generation. At the same time, it can only support fixed lengths, discarding those shorter than the fixed length, and splitting longer video clips, which is not conducive to training the model for video dubbing from videos with natural duration distributions. 

To address these challenges, we propose Kling-Foley, a large-scale multimodal V2A generation model that generates high-quality audio synchronized with video content. Our method is based on multimodal diffusion transformer, using large-scale variable-length multimodal data, combining video semantic alignment, audio-visual synchronization alignment, and text conditions to achieve accurate joint control modeling of multi-scenario sound effects.

Critically, sound effect generation tasks lack an available multimodal benchmark that includes visual, auditory, and caption information. Most existing benchmarks support only partial modalities. For example, AudioSet~\citep{gemmeke2017audio} provides audio with category labels, while VGGSound~\citep{chen2020vggsound} includes visual, audio, and label information. These datasets are not sufficient for a comprehensive evaluation of sound generation models.
To address this gap, we construct a high-quality and reliable benchmark dataset named Kling-Audio-Eval, based on a self-developed labeling system and a rigorous manual annotation process. To the best of our knowledge, Kling-Audio-Eval is the first benchmark for sound effect generation that includes video, video caption, audio, audio caption, and sound event labels.
The dataset contains 20,935 manually annotated samples and covers nine major sound event scenarios, such as traffic, human sounds, animal sounds, and more. 

In addition, we evaluate model performance from multiple perspectives by conducting assessments across four dimensions: Distribution Matching, Audio Quality, Semantic Alignment, and Temporal Alignment. This multi-dimensional approach provides a comprehensive understanding of the model's capabilities.

Our key contributions are as follows:
\begin{itemize}
    \item We introduce Kling-Foley, a novel V2A framework that generates high-fidelity audio perfectly synchronized with video content, achieving SOTA performance across audio quality, semantic alignment, and audiovisual synchronization metrics.
    
    \item We propose a visual semantic representation module and an audio-visual synchronization module that align video features with audio representations at each frame, enhancing semantic relevance and temporal coherence alongside text conditioning.
    
    \item We design a universal latent audio codec enabling high-fidelity modeling of various scenarios, including sound effect, speech, singing, and music.

    
    \item We release Kling-Audio-Eval, the first industry-scale multimodal benchmark with synchronized video, text descriptions, and audio featuring 20,935 manually annotated samples across nine sound scenarios, addressing the limitations of existing datasets through comprehensive multimodal annotations and scenario coverage.
\end{itemize}

\section{Related Work}

\subsection{Text-to-Audio}

Early approaches to audio generation leveraged GANs \cite{Dong2017MuseGANMS}, normalizing flows \cite{Kong2020HiFiGANGA}, and VAEs \cite{Oord2017NeuralDR}. Recent models largely follow two architectural paradigms: Transformer-based and Diffusion-based.

Transformer-based models like AudioGen \cite{Kreuk2022AudioGenTG} use autoregressive transformer to predict discrete audio tokens, while MAGNET \cite{Ziv2024MaskedAG} introduces masked generative modeling for non-autoregressive generation.

Diffusion-based methods such as DiffSound \cite{Yang2022DiffsoundDD} decode text into mel-spectrogram tokens via diffusion. Latent diffusion techniques \cite{Rombach2021HighResolutionIS} further drive models like AudioLDM \cite{Liu2023AudioLDMTG}, AudioLDM2 \cite{Liu2023AudioLDM2L}, Tango \cite{Ghosal2023TexttoAudioGU}, and Make-An-Audio \cite{Huang2023MakeAnAudioTG}. These models commonly rely on shared audio-text embedding spaces (e.g., CLAP \cite{Elizalde2023CLAPLA}) and spectrogram autoencoders, with strategies such as instruction tuning \cite{Chung2022ScalingIL} or pseudo prompt enhancement to improve data efficiency.

\subsection{Video-to-Audio}

V2A generation aims to synthesize sound from silent videos. Early work like \cite{Zhou2017VisualTS} used SampleRNN \cite{Mehri2016SampleRNNAU} to directly generate waveforms from frames. Later methods, such as SpecVQGAN \cite{Iashin2021TamingVG} and Im2Wav \cite{Sheffer2022IHY}, use visual features (e.g., CLIP \cite{Radford2021LearningTV}) to condition transformer models. Diff-Foley \cite{Luo2023DiffFoleySV} enhances temporal alignment via large-scale audio-visual pretraining and latent diffusion.

However, audio-visual datasets like VGGSound \cite{Chen2020VggsoundAL} are relatively small (550 hours), limiting model scalability. To address this, recent works extend pretrained TTA models for video input \cite{Jeong2024ReadWA, Mo2024TexttoAudioGS, Xing2024SeeingAH, Wang2023V2AMapperAL, Xie2024SonicVP, Zhang2024FoleyCrafterBS}. For example, Seeing and Hearing \cite{Xing2024SeeingAH} uses ImageBind \cite{Girdhar2023ImageBindOE} to convert video into text for AudioLDM, while V2A-Mapper \cite{Wang2023V2AMapperAL} maps visual features to CLAP embeddings. To incorporate temporal dynamics, models like SonicVLM \cite{Xie2024SonicVP}, ReWaS \cite{Jeong2024ReadWA}, and FoleyCrafter \cite{foleycrafter} integrate time-aware control modules.

Recent advancements in V2A generation have emphasized multimodal alignment and cross-modal contrastive learning to improve semantic relevance and temporal synchronization. VATT (Video-and-Text-to-Audio Transformer) \cite{Akbari2021VATTTF} introduces a convolution-free Transformer architecture that jointly learns representations from video, audio, and text modalities. 
In parallel, VTA-LDM (Video-to-Audio Latent Diffusion Model) \cite{Xu2024VideotoAudioGW} employs a CLIP-based vision encoder to extract high-resolution visual features and integrates auxiliary embeddings (e.g., positional embeddings, textual prompts) to guide audio generation. 
V-AURA \cite{Viertola2024TemporallyAA} introduces a cross-modal feature fusion strategy in an autoregressive framework, leveraging a high-framerate visual feature extractor (6× higher than prior work) to capture fine-grained motion details.
For efficiency, FRIEREN \cite{Wang2024FrierenEV} employs rectified flow matching (RFM) with reflow and one-step distillation. 
MMAudio \cite{mmaudio} proposes a joint training paradigm combining audio-visual (VGGSound) and audio-text (WavCaps) data within a unified multimodal transformer.

\subsection{Audio Representation}

The masked-based self-supervised learning paradigm, validated in natural language processing (NLP) with BERT \cite{devlin2019bert} and in computer vision with MAE \cite{he2022masked}, has become a mainstream approach for audio representation learning. Most models in this domain operate on audio spectrograms. Early works such as SSAST \cite{gong2022ssast} pioneered the framework of applying masked modeling to spectrogram patches. This approach was further developed by models like Audio-MAE \cite{huang2022masked} and MaskSpec \cite{chong2023masked}, which are heavily inspired by MAE and employ high-ratio random masking with an asymmetric encoder-decoder architecture to reconstruct original spectral features. 

Current reconstruction-based audio representation learning methods can be divided into discrete and continuous types. Discrete representations, commonly produced via residual vector quantization, compress audio into discrete tokens and reconstruct them with a decoder. Models such as EnCodec \cite{defossezhigh}, DAC \cite{kumar2023high} and AudioDec \cite{wu2023audiodec} follow this paradigm. Continuous representations do not rely on quantization. Instead, they learn smooth latent spaces through models like AudioLDM \cite{liu2023audioldm,liu2024audioldm}, Tango \cite{ghosal2023text,majumder2024tango}, Make-An-Audio \cite{huang2023make1, huang2023make}, and DiffRhythm \cite{ning2025diffrhythm}, which use variational encoders to capture audio features. 

Contrastive learning has become a pivotal method in cross-modal representation learning, with vision models like CLIP~\cite{radford2021learning}, Florence~\cite{yuan2021florence}, and ALIGN~\cite{jia2021scaling} effectively aligning image and text in a shared semantic space. In the audio domain, similar frameworks such as Wav2CLIP~\cite{wu2022wav2clip}, AudioCLIP~\cite{guzhov2022audioclip}, and CLAP~\cite{wu2023large} have achieved strong performance by learning global semantic representations of audio.

\section{Preliminary}

\subsection{Conditioning Encoders}
\label{main conditioner}
    Building upon the original SD3 framework \cite{sd3}, we introduce several targeted improvements to enhance both semantic representation and temporal alignment across modalities for multimodal audio generation: 

    \textbf{Text Encoder.} We adopt the T5-Base model proposed by Raffel et al.~\cite{t5-base} as the backbone for textual representation. T5 (Text-to-Text Transfer Transformer) introduces a unified framework that reformulates all NLP tasks into a text-to-text paradigm, allowing for a highly flexible and task-agnostic approach to model training and inference. Pretrained on the Colossal Clean Crawled Corpus (C4), T5-Base benefits from exposure to vast and diverse language data, enabling it to capture nuanced semantic relationships and contextual dependencies across a broad range of domains. In our system, the T5-Base functions as the primary text encoder, transforming natural language inputs—such as prompts, descriptions, or queries—into rich latent embeddings. These embeddings serve as a semantic anchor that guides downstream multimodal alignment and generation processes. Furthermore, the model’s ability to generalize well across tasks allows us to maintain high adaptability and performance with minimal task-specific tuning.

    \textbf{Vision Encoder.} For the extraction of high-level visual semantics and effective multimodal grounding, we incorporate the ViT-bigG-14-QuickGELU model within the MetaCLIP framework, as introduced by Ma et al.~\cite{meta-clip}. ViT-bigG, a large Vision Transformer architecture, utilizes QuickGELU activations and benefits from extensive pretraining on large-scale, high-quality image-text pairs. The MetaCLIP training strategy enhances this foundation by employing an expert clustering mechanism to better capture the alignment between visual and linguistic modalities. This results in a visual encoder that excels at producing domain-robust, semantically meaningful image embeddings. These embeddings not only preserve fine-grained visual details but also remain aligned with their textual counterparts, making the model particularly suitable for tasks involving vision-language fusion, retrieval, and generation. In our multimodal pipeline, the visual encoder plays a critical role in grounding generated content in the visual domain, ensuring that the output remains coherent and contextually relevant.    
    
    \textbf{Align Encoder.} Temporal consistency across modalities is crucial in many real-world multimodal applications, such as video generation, dubbing, or speech-driven animation. To address this, we utilize the Synchformer model proposed by Iashin et al.~\cite{synchformer}, a transformer-based architecture explicitly designed for audio-visual synchronization. Synchformer leverages sparse synchronization cues—like lip movements, phoneme timing, and audio features - to infer fine-grained alignment across the temporal dimension. Unlike traditional alignment methods that may rely on dense supervision or heuristic rules, Synchformer employs self-attention and cross-modal transformer to model temporal dependencies efficiently. Its compact yet expressive design enables it to operate robustly in both constrained and unconstrained environments, ensuring that multimodal outputs (e.g., lip-synced avatars or synchronized video narration) maintain a high level of realism and temporal coherence. In our system, the align encoder acts as a mediator that refines and aligns latent representations from the text and vision encoders, ensuring consistent temporal flow and cross-modal harmony throughout the generative process.
    
\subsection{Multimodal Diffusion Transformer}
To enable effective generation from arbitrary combinations of text, video, and audio inputs, we design a unified multimodal conditioning framework. This framework supports flexible pairwise or tri-modal combinations such as TTA, V2A, and Text-Video-to-Audio (TV2A) in a single model. At the core of our design is a modular encoder architecture and a joint conditioning mechanism that harmonizes modality-specific representations into a temporally aligned and semantically coherent latent space.

We build upon the MM-DiT framework introduced in SD3~\cite{sd3}, extending it with enhanced temporal synchronization modules and a dynamic masking strategy. During training, each modality is encoded independently, and missing modalities are replaced with learnable placeholders. The encoded features are then projected into a shared embedding space, where audio and video tokens are augmented with RoPE-based temporal positional encodings~\cite{su2024roformer} to facilitate alignment. To enable generation over sequences of variable durations, we further introduce learnable duration embeddings as timing-aware vectors, which are fused with global conditioning features. These fused embeddings serve as the foundation for downstream joint attention and generation. Additionally, inspired by FLUX\cite{flux2024}, we introduce audio-only single-modality blocks by simply removing the data streams of the other two modalities (i.e., converting joint attention into self-attention).

\subsection{Flow Matching}
\label{flow matching}
During training, we use a conditional flow matching objective \cite{lipman2022flow} for generative modeling \cite{tong2023improving}. This modeling method utilizes conditions $C$ such as text or video embedding after encoding to learn a time-dependent conditional velocity vector field function $v_\theta(t,C,x)$ that describes the direction and speed of the input $x$ at timestep $t$, the variable $\theta$ represents the network parameters that need to be learned. This function guides the Gaussian noise latent variable $x_0$ to fit and approximate the target latent audio variable $x_1$ using an ordinary differential equation solver to numerically integrate over the time interval $ t\in[0, 1] $. The function $u$ represents the target conditional vector field, and $p$ represents the conditional probability path. $q$ denotes the distribution of training data.
\begin{equation}
\label{e:cfm}
    \mathcal{L}_{CFM}(\theta) = \mathbb{E}_{t,q,p} \big\| v_\theta(t, C, tx_1+(1-t)x_0) - u(t, tx_1+(1-t)x_0) \big\|^2,
\end{equation}

During inference, we set the time step $t$ to 0.05 and employ the Euler method to numerically integrate $v_\theta(t,C,x)$, reconstructing the final latent audio representation from the initial Gaussian noise.

\subsection{Aligned RoPE Positional Embedding}
Precise temporal alignment plays a critical role in effectively synchronizing audiovisual content. To incorporate temporal information into the attention layers, we employ Rotary Positional Embeddings (RoPE)\cite{su2024roformer}, which are applied to the queries and keys within the visual and auditory branches prior to joint attention (see Figure \ref{fig:V2A}). Since the textual modality does not possess an intrinsic temporal structure comparable to audio and video, it is excluded from this positional encoding step. Furthermore, because visual tokens are sampled at a lower temporal resolution than audio tokens, we proportionally scale the frequency components of the visual positional embeddings to match the higher temporal rate of the auditory modality. This adjustment facilitates alignment across modalities. While these modified embeddings mitigate temporal misalignment, they are insufficient on their own to guarantee robust synchronization. Therefore, we introduce a dedicated synchronization module to further improve temporal coherence.

\section{Kling-Foley}
\subsection{Overview}
Inspired by MMAudio\cite{mmaudio} we propose Kling-Foley. The core of our approach is to model the interactions between video, audio, and text modalities. We adopt the MM-DiT block design of SD3\cite{sd3} and introduce two new components for temporal alignment: aligned RoPE\cite{su2024roformer} position embeddings to adapt to sequences of different frame rates, and 1D convolutions and MLPs to capture local temporal structure. At the same time, a duration learnable module is added to control the production of variable-length audio from naturally distributed video duration features. In addition, we incorporate audio-specific unimodal blocks based on FLUX \cite{flux2024} to make the network deeper with the same parameters without sacrificing multimodal capabilities. This architecture allows the model to selectively focus on different modalities based on the input, supporting joint training of audio-visual and audio-text data.

\begin{figure*}[t]
 \centering
 \includegraphics[width=\linewidth]{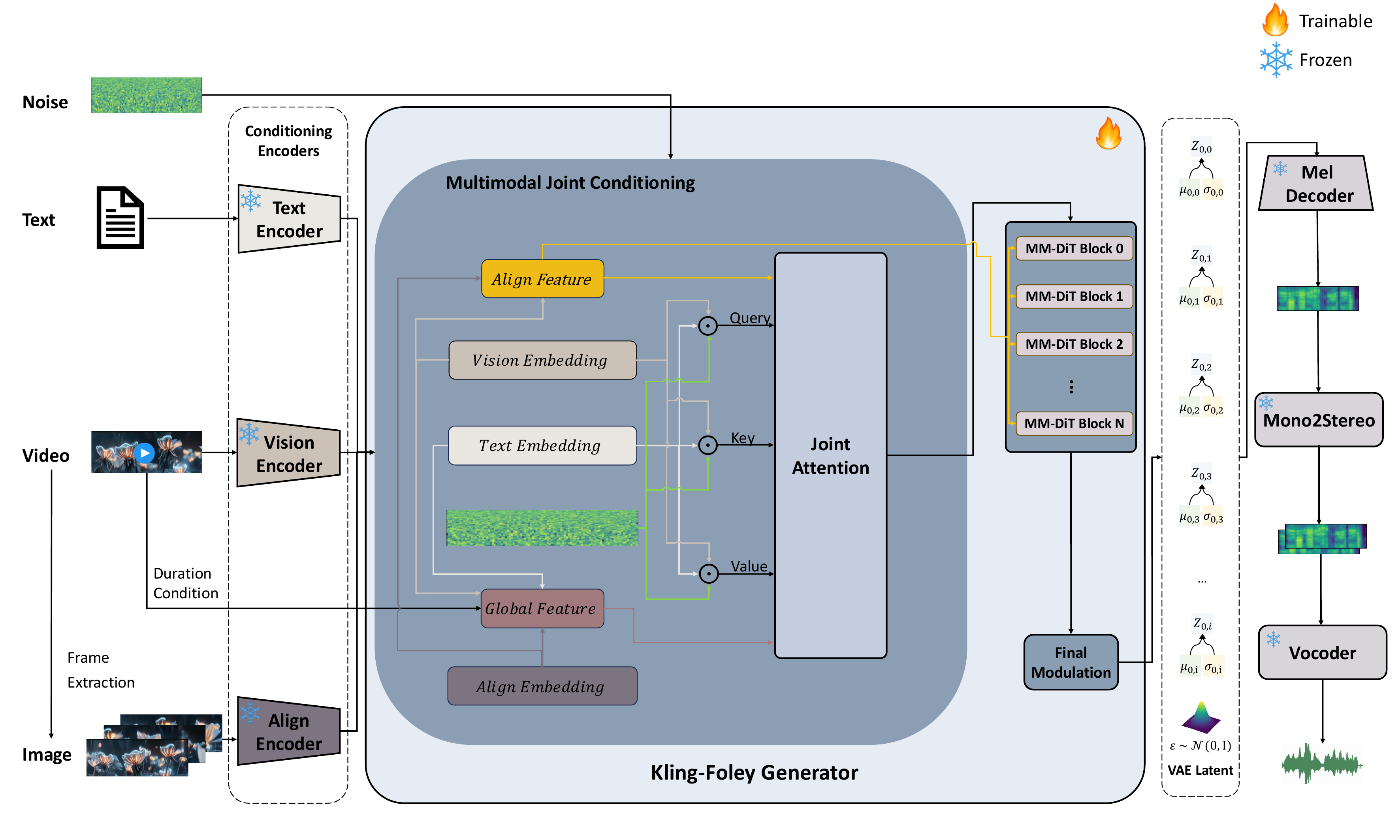}
 \caption{The core of Kling-Foley is a multimodal-controlled flowmatching model. Text, video, and temporally extracted video frames serve as conditional inputs. The multimodal features are then fused via a Multimodal Joint Conditioning module, which feeds into the MMDit Block for processing. This module predicts VAE latents, which a pretrained mel decoder subsequently reconstructs into a monaural mel-spectrogram. The monaural spectrogram is then converted to stereo spectrogram via a Mono2Stereo module. Finally, the stereo spectrogram is passed through a vocoder to generate the output waveform.}
 \label{fig:V2A}
\end{figure*}

\subsection{Variable Duration Control}
To support variable-length audio-visual generation and enhance temporal control, we introduce discrete duration embeddings as part of the global conditioning mechanism \cite{stableaudio}. Specifically, two scalar properties are computed per training clip: the start time seconds-start and the total duration seconds-total of the original video-audio sequence. These values are embedded into learnable per-second embeddings and concatenated with global textual and visual features.
The resulting timing-aware global condition is fused with the flow timestep embedding via a shallow MLP and applied to all transformer layers using Adaptive Layer Normalization \cite{tango}. Each adaLN layer modulates token-wise activations by scaling and shifting normalized features based on the global conditioning vector.

\subsection{Joint Attention Mechanism}
In our architecture, we enable cross-modal communication through a joint attention strategy (see Figure \ref{fig:V2A}). Inspired by prior work\cite{esser2024scaling}, we integrate the query, key, and value matrices across the textual, visual, and auditory modalities into a unified attention computation. Specifically, these modality-specific components are concatenated and passed through a shared scaled dot-product attention module\cite{vaswani2017attention}. After the attention operation, the resulting unified output is segmented back into the original three modalities based on the initial token groupings. While this joint mechanism facilitates rich cross-modal interaction, we emphasize that, on its own, it does not ensure temporal synchrony across streams such as audio and video.

\begin{algorithm}[t]
\caption{Multimodal Conditional Training Strategy}
\label{alg:modality_training}
\begin{algorithmic}
\small
\STATE \textbf{Input:} Audio $x$, Video $V$ (optional), Text $T$ (optional)
\STATE \textbf{Hyperparameters:} total steps $S$, frame rates, number of transformer blocks $N_1$, $N_2$
\STATE \textbf{Initialize:} Learned empty embeddings $e_v$, $e_t$ for missing V/T

\FOR{each training step $s=1$ to $S$}
    \STATE \textbf{Stage 1: Modality-Specific Encoding}
    \IF{Video $V$ available}
        \STATE Extract sync features $F_{\text{sync}} \leftarrow \text{SyncFormer}(V)$
        \STATE Extract video features $F_v \leftarrow \text{MetaCLIP}_{\text{visual}}(V)$
    \ELSE
        \STATE $F_{\text{sync}} \leftarrow e_v$, $F_v \leftarrow e_v$
    \ENDIF

    \IF{Text $T$ available}
        \STATE $F_t \leftarrow \text{T5}(T)$
    \ELSE
        \STATE $F_t \leftarrow e_t$
    \ENDIF

    \STATE Encode audio $x$ to latent representation $x_0$ with diffusion noise

    \STATE \textbf{Stage 2: Conditional Synchronization Module}
    \STATE Project $F_{\text{sync}}$ and upsample to frame-aligned sync feature
    \STATE Generate align feature $a_f$ and global feature $g_f$ 

    \STATE \textbf{Stage 3: Transformer-based Joint Processing}
    \STATE Project and align $F_v$, $F_t$, $x_0$
    \STATE Inject positional embeddings (RoPE) to visual/audio queries and keys

    \FOR{$i = 1$ to $N_1$}
        \STATE Apply multimodal transformer block with joint attention over $F_v$, $F_t$, $x_0$ and conditions $a_f$, $g_f$
    \ENDFOR

    \FOR{$i = 1$ to $N_2$}
        \STATE Apply single-modal transformer block to refine audio flow path
    \ENDFOR

    \STATE \textbf{Stage 4: Output Flow Prediction}
    \STATE Apply adaptive LayerNorm and 1D-Conv to predict audio flow $w$
    \STATE Use $w$ in reverse diffusion to reconstruct waveform

    \STATE \textbf{Stage 5: Loss Computation and Optimization}
    \STATE Compute losses: $\mathcal{L}_{CFM}(\theta) = \mathbb{E}_{t,q(x_1),p_t(x\vert x_1)} \big\| v_t(x) - u_t(x\vert x_1) \big\|^2$    
    \STATE Backpropagate and update model
\ENDFOR
\end{algorithmic}
\end{algorithm}

\subsection{Flexible Pairwise Training}
To effectively support multimodal generation under arbitrary combinations of available inputs (text, video, and audio), we adopt a conditional training strategy that reflects the modular structure outlined in Algorithm~\ref{alg:modality_training}. Each modality is first encoded independently: video inputs are processed by MetaCLIP and SyncFormer to extract semantic and synchronization features, while textual inputs are encoded via T5. Missing modalities are substituted with learned placeholder embeddings ($e_v$, $e_t$), ensuring a consistent representation space. 

The synchronization features are projected and upsampled to produce two types of conditioning tokens: align feature and global feature, the latter incorporating learnable embeddings of start time and total duration. These global tokens are fused with diffusion timestep embeddings via a shallow MLP and modulate each transformer layer through Adaptive LayerNorm~\cite{tango}.

In the joint transformer stage, modality-specific features are projected into a shared latent space. RoPE embeddings~\cite{su2024roformer} are added to audio and visual tokens—rescaled for temporal alignment—to encode temporal structure. Joint attention enables cross-modal interaction, while audio-only transformer blocks~\cite{flux2024}, applied after joint fusion, provide efficient unimodal refinement, benefiting tasks like audio continuation and TTA.

\subsection{Latent Audio Codec}

The latent audio codec extends our prior VQ-CTAP framework\cite{qiang2025vq,qiang2024learning}, inheriting its core components while introducing key modifications to optimize audio reconstruction.
As illustrated in Figure \ref{fig:VAE}, the core of the latent audio codec is a Mel-VAE composed of three primary components: a mel encoder, a mel decoder, and a discriminator. The audio encoder processes an input waveform sampled at 44.1 kHz, generating embeddings at a rate of 43 Hz (equivalent to 1024 times downsampling relative to the input sampling rate). Critically, the VAE structure enables the model to learn a continuous and complete distribution of latent spaces, significantly enhancing audio representation capabilities.

\subsubsection{Structure}

\begin{figure}[t]
 \centering
 \includegraphics[width=\linewidth]{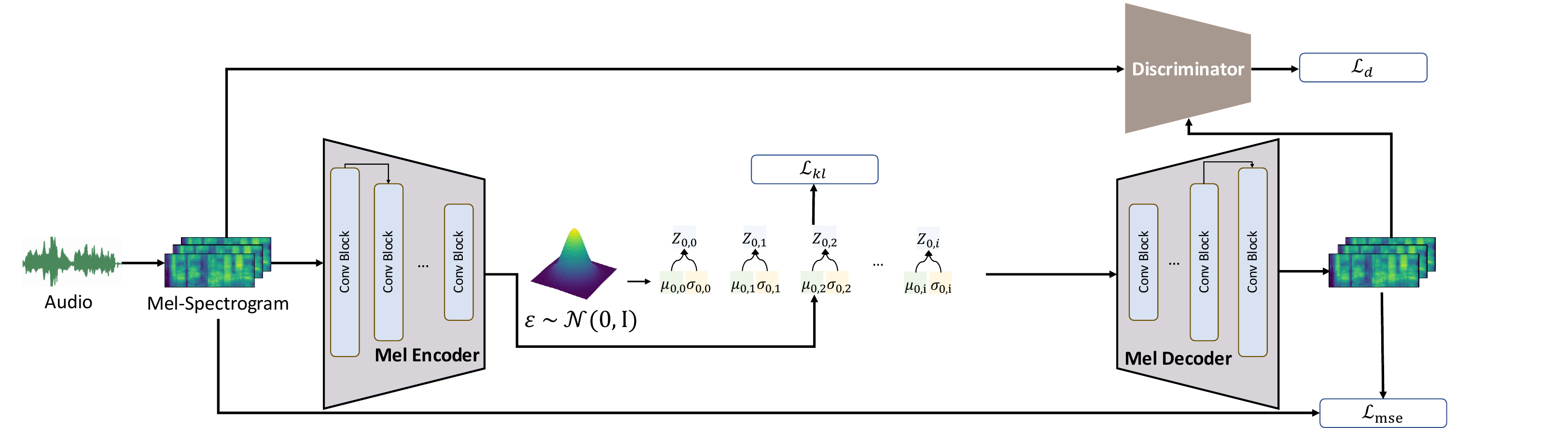}
 \caption{The main body of latent audio codec is a Mel-VAE, which jointly trains a mel encoder, a mel decoder, and a discriminator. The VAE structure enables the model to learn a continuous and complete distribution of latent spaces, significantly enhancing its audio representation capabilities.}
 \label{fig:VAE}
\end{figure}

As illustrated in Figure \ref{fig:VAE}, $A_{in}$ denotes the input batch of target audio data, $A_{in} \in \mathbb{R}^{B \times T_s \times D_s}$, where $B$ is the batch size, $T_s$ is the number of time frames, and $D_s$ is the number of spectral components (mel-spectrogram bands). The $S_{in}$ is passed through the audio encoder: $A = \mathrm{AudioEncoder}(A_{in})$

where ${A} \in \mathbb{R}^{B \times T_s/2 \times d}$ are the target audio representations. The audio encoder compress the length of the audio representations by a factor of 2.

Following encoding, the model utilizes the VAE structure\cite{qiang2022style} to parameterize the latent distribution.

In Equation (5), $D_{KL}$ refers to the KL loss, $\mathcal{N}(\cdot)$ represents a Gaussian distribution, and $({\hat{\mu}}, {\hat{\sigma}})$ denotes the $(mean, variance)$ of the audio representation latent space distribution. Random operations in the network cannot be processed by backpropagation, "reparameterization trick" is introduced to VAE: $\boldsymbol{z} = {\hat{\mu}} + {\hat{\sigma}} \odot \phi ; \phi \sim \mathcal{N}(0, I)$

To address the KL collapse problem\cite{qiang2023improving}, a margin $\Delta$ is introduced to limit the minimum value of the KL loss as shown: 

\begin{equation}
\begin{aligned}
 \mathcal{L}_{kl} = max(0, D_{KL}[\mathcal{N}({\hat{\mu}},{\hat{\sigma}}^2)||\mathcal{N}(0, I)]-\Delta)
\end{aligned}
\end{equation}

To enable reconstruction capability using the pre-trained latent representations, the latent variable $\boldsymbol{z}$ serves as input to the audio decoder for predicting the mel-spectrogram: $A_{d} = \mathrm{AudioDecoder}(\boldsymbol{z})$

A mean squared error (MSE) loss is used to compare the predicted mel-spectrogram $A_{d}$ with the ground-truth mel-spectrogram $A_{in}$:

\begin{equation}
\begin{aligned}
\mathcal{L}_{mse} = MSE(A_{in}, A_{d})
\end{aligned}
\end{equation}

The discriminator module is pivotal for adversarial training, enforcing the generator to produce high-fidelity audio spectrograms indistinguishable from real data. Its design integrates multi-scale feature analysis, gradient regularization, and dynamic loss weighting, addressing spectral artifacts and training instability common in audio synthesis. 

The generator aims to deceive the discriminator by maximizing the probability of generated spectrograms being classified as real. This is formalized as a non-saturating loss: 

\begin{equation}
\begin{aligned}
\mathcal{L}_g = -\mathbb{E}\big[D\big(A_d\big)\big]
\end{aligned}
\end{equation}

Where $D(\cdot)$ represents the discriminator's output logit.

The discriminator learns to distinguish real and generated spectrograms via a weighted binary objective:

\begin{equation}
\begin{aligned}
\mathcal{L}_d = \underbrace{\gamma_{step} \cdot \mathcal{L}_{adv}}_{\text{adversarial loss}} + \underbrace{\lambda_{R1} \cdot \mathcal{R}_1}_{\text{gradient penalty}}
\end{aligned}
\end{equation}

\begin{equation}
\begin{aligned}
\mathcal{L}_{adv} = \mathbb{E}[\max(0,1 - A_{in})] + \mathbb{E}[\max(0,1 + D(A_{d}))]
\end{aligned}
\end{equation}

\begin{equation}
\begin{aligned}
\mathcal{R}_1 = \mathbb{E}_{x}\big[|\nabla_x \mathrm{AudioDecoder}(\boldsymbol{z}) |^2_2\big]
\end{aligned}
\end{equation}

Where $\gamma_{step}$ and $\lambda_{R1}$ represent the dynamic weighting.

In the experiment, we employ a mel-spectrogram encoder structurally similar to that used in Make-An-Audio2 \cite{huang2023make}. This encoder comprises 32 stacked 1D-convolutional layers. The mel decoder mirrors this architecture with transposed convolutions for mel-spectrogram reconstruction.

\subsubsection{Multi-Stage Stepping Optimization Strategy}

A stepping optimization strategy is designed to ensure effective model convergence by gradually injecting and adjusting the influence of various loss components, as shown in Algorithm \ref{alg:training}. The training process involves the following losses: $\mathcal{L}_{kl}$, $\mathcal{L}_{mse}$, $\mathcal{L}_{d}$, and $\mathcal{L}_{g}$. The variable $step$ represents the current training step.
Initially, the model is trained using $\mathcal{L}_{mse}$. When the $step$ exceeds the specified starting step for $\mathcal{L}_{kl}$, $\mathcal{L}_{kl}$ is added to the training process. The weight for $\mathcal{L}_{kl}$ increases gradually as the training progresses. Once the $step$ surpasses the specified ending step, the weight for $\mathcal{L}_{kl}$ is fixed at $kl\_upper$.
Similarly, when the $step$ exceeds the specified starting step for $\mathcal{L}_{d} \& \mathcal{L}_{g}$, These losses are incorporated into the training process. The weight for $\mathcal{L}_{d} \& \mathcal{L}_{g}$  also increases gradually during training. Once the $step$ exceeds the specified ending step, the weight for $\mathcal{L}_{d} \& \mathcal{L}_{g}$ is fixed at $gan\_upper$.
In the final training stage, we freeze both the audio encoder and VAE, training exclusively the audio decoder and discriminator. This focused optimization addresses potential error propagation by fine-tuning the vocoder using mel-spectrograms generated from the audio decoder's output.
Algorithm \ref{alg:training} outlines the step-wise inclusion of different losses and their corresponding weight adjustments. This optimization strategy aims to facilitate effective model convergence by gradually introducing and adjusting the influence of various loss components throughout the training process.

\begin{algorithm}[h]
\caption{Multi-Stage Stepping Optimization Strategy}\label{alg:training}
\begin{algorithmic}
\small
\STATE Initialize hyperparameters: 
\STATE \hskip1.5em $stage1_{end}$, $stage2_{end}$, $stage3_{end}$ \COMMENT{Phase transition steps}
\STATE \hskip1.5em $kl_{upper}$, $gan_{upper}$ \COMMENT{Max loss weights}
\FOR{each training $step$}
    \IF{$step \leq stage1_{end}$} 
        \STATE \textbf{Stage 1: Base Reconstruction} 
        \STATE Train full model with $\mathcal{L}_{mse}$ \COMMENT{Pure MSE focus}
    \ELSIF{$step \leq stage2_{end}$}
        \STATE \textbf{Stage 2: Regularization Enhancement} 
        \STATE $\gamma \gets kl_{upper} \cdot \min(1, \frac{step - stage1_{end}}{stage2_{end} - stage1_{end}})$
        \STATE $\mathcal{L}_{total} \gets \mathcal{L}_{mse} + \gamma \mathcal{L}_{kl}$
        \STATE Train full model with $\mathcal{L}_{total}$ 
    \ELSIF{$step \leq stage3_{end}$}
        \STATE \textbf{Stage 3: Adversarial Introduction} 
        \STATE $\delta \gets gan_{upper} \cdot \min(1, \frac{step - stage2_{end}}{stage3_{end} - stage2_{end}})$
        \STATE $\mathcal{L}_{total} \gets \mathcal{L}_{mse} + kl_{upper}\mathcal{L}_{kl} + \delta(\mathcal{L}_{g} + \mathcal{L}_{d})$
        \STATE Train full model with $\mathcal{L}_{total}$
    \ELSE 
        \STATE \textbf{Stage 4: Decoder Refinement} 
        \STATE $\mathcal{L}_{total} \gets \mathcal{L}_{mse} + kl_{upper}\mathcal{L}_{kl} + gan_{upper}(\mathcal{L}_{g} + \mathcal{L}_{d})$
        \STATE Freeze audio encoder and VAE parameters
        \STATE Train only audio decoder and discriminator with $\mathcal{L}_{total}$
    \ENDIF
\ENDFOR
\end{algorithmic}
\end{algorithm}

\subsection{Mono-to-Stereo}
The process utilizes a Mono2Stereo module to convert the monaural mel-spectrogram into dual-channel mel-spectrograms. Critically, this module only predicts the ratio of the left and right mel-spectrograms relative to the monaural mel-spectrogram. This targeted prediction significantly reduces data dependency and enhances training stability. Finally, these left and right mel-spectrograms are processed by a vocoder to generate corresponding waveforms for each channel, which are concatenated along the channel dimension to produce the final stereo audio.

\section{Data}
\subsection{Overview} \label{Overview}
To train a multimodal generative model capable of synthesizing diverse and realistic sound effects, it is essential to construct a large-scale training dataset that is broad in coverage and tightly aligned across modalities.
Current research in sound effect generation faces two critical limitations. First, most existing datasets are relatively small, typically containing only tens of thousands of audio samples, which are insufficient to support the training of large-scale generative models that require high data diversity. 
Second, the majority of these datasets are incomplete in modality structure—lacking alignment among audio, video, and natural language—which significantly limits the model’s ability to utilize conditional inputs effectively.
For instance, the VGG-Sound dataset~\cite{chen2020vggsound} contains audio-video pairs but only includes coarse class labels, without natural language descriptions of the sound content. On the other hand, datasets such as AudioCaps~\cite{kim2019audiocaps}, Clotho~\cite{drossos2020clotho}, and WavCaps~\cite{mei2024wavcaps} focus primarily on audio-caption alignment, yet do not include accompanying video streams, making them less suitable for training generation models conditioned on both vision and language.

To address these challenges, we construct a new large-scale multimodal sound effects dataset from scratch, consisting of over 100 million samples. Each sample contains a raw video segment, a corresponding monaural audio clip, and a structured textual description of the audio. 
The three modalities are tightly aligned and sourced from real-world online video content.

\begin{figure}[H]
 \centering
 \includegraphics[width=\linewidth]{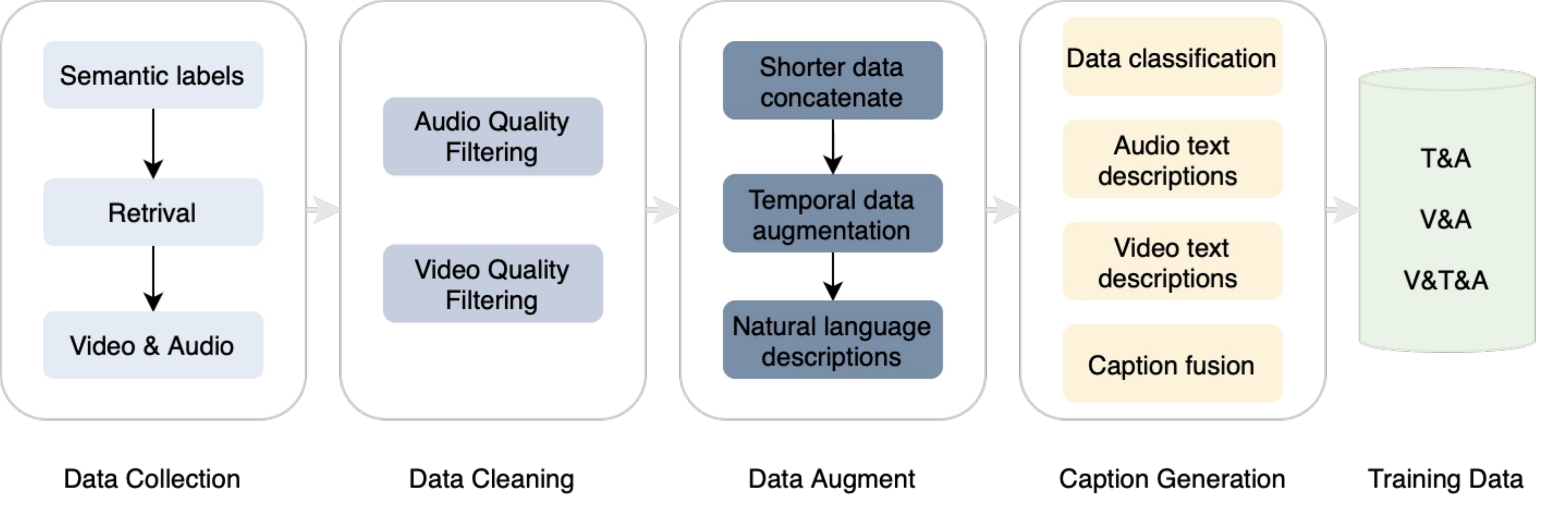}
 \caption{Audio and video data undergo preprocessing and quality filtering to obtain high-quality single-event audio and video segments. Subsequently, synthetic multi-event audio samples are generated through temporal augmentation, and large models are used to generate and extract keywords and classification captions for audio and video. Finally, various caption information is combined to produce the final training captions.}
 \label{fig:DP}
\end{figure}

\subsection{Data Construction} \label{Data Construction}

In this work, we constructed three types of paired data: text-audio, video-audio, and video-text-audio. Our overall data processing workflow is shown in Figure~\ref{fig:DP}.

\textbf{Data Collection} 
The generative capacity of sound synthesis models is largely determined by the variety of sound sources and the range of semantic labels present in the training data. To ensure broad coverage, we build our label set based on the hierarchical structure defined in the AudioSet~\cite{gemmeke2017audio} ontology, selecting categories from its top three levels. This ontology provides a clear semantic hierarchy, serving as a principled foundation for constructing a systematic keyword vocabulary for data mining.

Using this label set, we derive a keyword bank to guide the large-scale retrieval process. We query video platforms using these keywords and filter candidate videos and channels based on metadata such as titles, descriptions, and tags to ensure semantic relevance.
To further enhance long-tail coverage and content diversity, we supplement the collected videos with manually curated resources and samples from existing open-source datasets. 
The resulting raw multimodal data serves as the foundation for our dataset and is subsequently passed through a dedicated cleaning pipeline to ensure quality and alignment across modalities.

\begin{figure}[t]
  \centering
  \includegraphics[width=\linewidth]{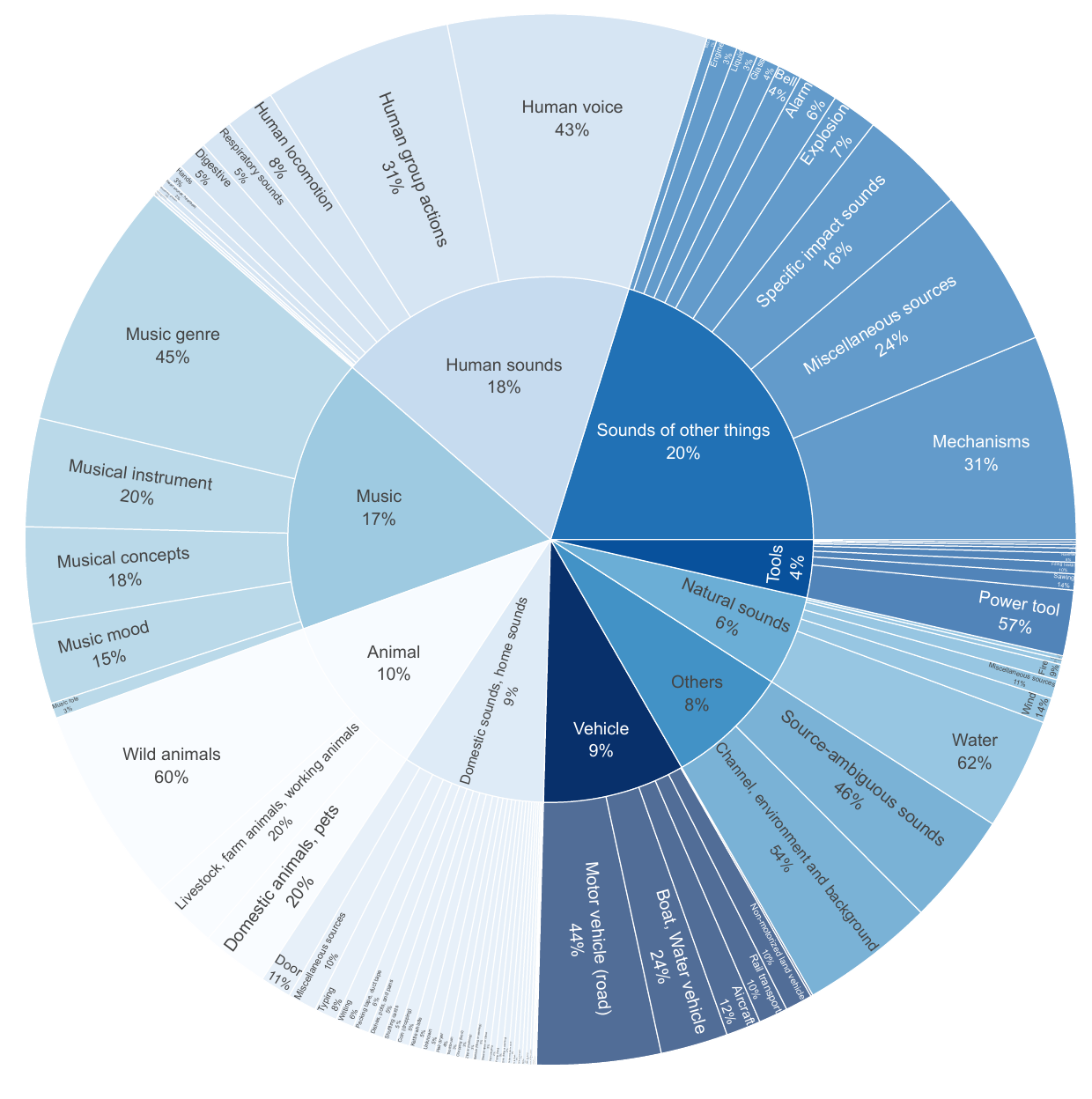}
  \caption{Category distribution of sound events in the training set. The broad coverage of real-world acoustic events  ensures the diversity and generalizability required for training open-domain sound generation models.}
  \label{fig:class_distribution}
\end{figure}
\textbf{Data Cleaning} 
We only retain data with video resolution above 720P and a small proportion of subtitles, and we uniformly convert the audio to WAV format with a 44k sample rate, 16-bit depth, and stereo channels. 
For audio, we perform quality filtering based on SNR, MOS score, clipping ratio, and audio bandwidth. We use VAD to select audio data with a silence ratio of less than 0.2. We employ the CLAP model to calculate the consistency between audio and text labels, retaining data with high consistency. 
Finally, we split longer videos and audio into 10-second segments.

\textbf{Data Augment} 
For shorter data, we concatenate shorter videos and audio to obtain data with a duration of 10 seconds, which enables the model to better response to dynamic visual input. 
To enhance the temporal alignment ability of our model, we introduce temporal data augmentation~\cite{huang2023make} by concatenating single-event video and audio clips according to different temporal rules to generate multi-event data. The text captions corresponding to the temporally augmented video and audio are obtained by merging the original single-event data text captions. 
Additionally, we extract key information such as sound sources, objects, scenes, emotions, gender, actions, and modifiers from the audio. Using a large model combined with the extracted keywords, we transform the unstructured original text descriptions into semantically complete natural language descriptions.

\textbf{Caption Extraction} 
Video and audio can typically yield textual descriptions containing different information. To obtain text captions that are as accurate, detailed, and complete as possible, we utilize both video and audio to derive the final text caption.
First, we use audio classification model to classify the video and audio, retaining data and corresponding category labels for four categories: sound effects, music, speech, and singing~\cite{ehtesham2025movie}. 
For different categories of data, we employ corresponding audio understanding large models to extract audio text descriptions from the audio, while also extracting video text descriptions from the video. 
Subsequently, we input the audio descriptions, video descriptions, and the enhanced natural language text descriptions into a large model to obtain the final fused text caption~\cite{polyak2024movie}.

\textbf{Training Data} 
As illustrated in Figure~\ref{fig:class_distribution}, we visualize the distribution of high-level sound categories in our training set. 
Our training data contains text\-audio, video\-audio, and video\-text\-audio three types of paired data. 
Our dataset spans a wide variety of real-world acoustic scenarios, including natural environments, human activities, animal sounds, mechanical operations, and transportation, providing a solid foundation for learning diverse generative patterns and improving the realism and controllability of synthesized audio.

\subsection{Benchmark Dataset}
Several audio-visual datasets have been proposed to support sound-related tasks, as shown in Table~\ref{tab:dataset_comparison}. AudioSet~\cite{gemmeke2017audio} is one of the largest general-purpose audio datasets, but its heavy reliance on human annotation leads to high construction costs. VGGSound~\cite{chen2020vggsound} improves scalability through audio-visual alignment, making it more practical for sound generation evaluation. EPIC-SOUNDS~\cite{huh2023epic} focuses on audio-driven actions, offering precise temporal boundaries and fine-grained labels.

However, a common limitation of these datasets is the lack of textual descriptions (captions) for both audio and video modalities, which hinders comprehensive evaluation in caption-aware or text-conditioned generation scenarios.
To address this issue, a common solution is to incorporate additional audio-text test sets, such as Clotho~\cite{drossos2020clotho} and AudioCaps~\cite{kim2019audiocaps}. 
While Clotho provides high-quality captions, its scale is limited. AudioCaps offers more annotated samples, but only covers audio modality and lacks video context. WavCaps~\cite{mei2024wavcaps}, though large-scale, is weakly labeled and unsuitable for evaluation.

\begin{table}[ht]
\centering
\caption{Comparison of our dataset with existing datasets. }
\label{tab:dataset_comparison}
\begin{tabular}{@{}lccccccc@{}}
\toprule
\textbf{Dataset}                & \textbf{\#Test clips}  &   \textbf{Length} & \textbf{Video} &\makecell{\textbf{Video}  \\ \textbf{Caption}} & \textbf{Audio} &\makecell{\textbf{Audio}  \\ \textbf{Caption}} &\makecell{\textbf{\#Class}}\\ \midrule
\makecell[l]{AudioSet ~\cite{gemmeke2017audio}}      & 18K     & 50h    & Yes   & No  & Yes  & No &527\\
\makecell[l]{VGGSound ~\cite{chen2020vggsound}}      & 15K     & 41.7h    & Yes & No  & Yes  & No &309\\
\makecell[l]{Epic Sounds ~\cite{huh2023epic}}      & 10K     & 13.9h    & Yes  & No  & Yes  & No &44\\
\makecell[l]{Clotho ~\cite{drossos2020clotho}}      & 1K     & 6h    & No  & No  & Yes   & Yes &-\\
\makecell[l]{AudioCaps ~\cite{kim2019audiocaps}}      & 1K     & 2.7h    & No  & No  & Yes  & Yes &-\\
\midrule 
\textbf{Kling-Audio-Eval}                 & \textbf{21K}  & \textbf{58.6h}   & Yes   & Yes  & Yes & Yes  & 1919 \\ \bottomrule
\end{tabular}
\end{table}

\begin{figure}[h]
  \centering
  \includegraphics[width=\linewidth]{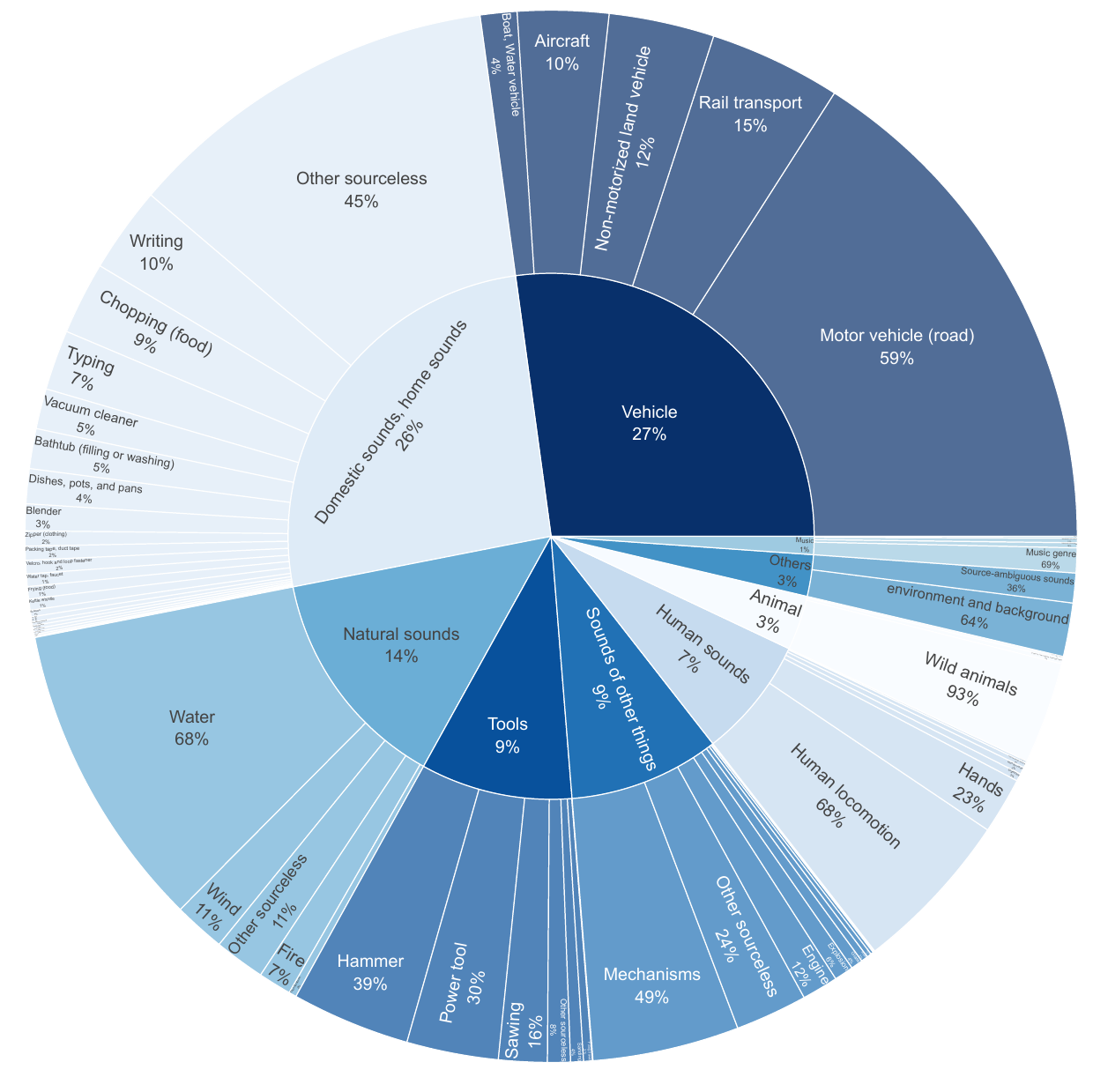}
  \caption{Distribution of various categories in the benchmark dataset.}
  \label{fig:benchmark_distribution}
\end{figure}

Despite these efforts, there remains no benchmark that jointly supports vision, audio, and language modalities for evaluating sound effect generation.
\textbf{To fill this gap, we introduce Kling-Audio-Eval — the first high-quality multimodal benchmark that combines video, video captions, audio, audio captions, and sound event labels}. Our dataset is constructed through a carefully designed taxonomy and extensive human annotation, enabling robust evaluation across multiple modalities and generation tasks.

Based on existing sound libraries and internal statistics, we selected the 1,000 most frequent third-level labels (Section~\ref{Overview}). These were chosen to ensure full coverage of all first- and second-level categories, providing broad content representation. Following a strict data cleaning process (Section~\ref{Data Construction}), we then selected 30,000 samples from these categories, each with pre-generated captions and sound event labels, for further human verification and annotation.

The manual annotation process covers four main aspects: audio captions, video captions, sound event labels, and audio-visual quality assessment, aiming to ensure consistency, accuracy, and practical usability. The specific annotation guidelines are as follows:
\begin{itemize}
    \item Caption Correction: Review the pre-generated audio and video captions, and revise any errors or omissions using concise and clear language.
    \item Modal Independence: Ensure that audio and video captions are annotated independently. For example, audio captions should not rely on visual information, and vice versa.
    \item Label Verification: Check whether the assigned first- and second-level labels match the actual content. If not, select the correct labels from the predefined taxonomy.
    \item Valid Sample Selection: Only retain audio-visual clips that meet the following criteria: (i) Foreground audio must not contain human voices. (ii) Sound effects must originate from tasks or objects, while speech, singing, and musical sounds must be produced by visible individuals in the video. (iii) Video duration must be at least 5 seconds, and the sound effect must last at least 2 seconds. (iv) Only ambient off-screen sounds are permitted, such as birdsong in a forest setting; (v) Videos are allowed to include watermarks, logos, or subtitles; (vi) Background music must not contain vocals.
\end{itemize}
In total, we collected 20,935 high-quality samples to form the final test set, and the original 1,000 third-level labels were further refined into 1,919. The category distribution is illustrated in Figure~\ref{fig:benchmark_distribution}.

\section{Experiments and Results}
\subsection{Experimental Settings}

We benchmark against the following methods:
V2A-Mapper \cite{Wang2023V2AMapperAL} maps CLIP visual embeddings into the CLAP audio-text space to enable AudioLDM-based generation, while FoleyCrafter \cite{foleycrafter} incorporates time-conditional estimators for improved temporal coherence.
VATT \cite{Akbari2021VATTTF} and VTA-LDM \cite{Xu2024VideotoAudioGW} adopt transformer-based architectures for joint video, audio, and text representation learning. 
V-AURA \cite{Viertola2024TemporallyAA} employs a high-framerate visual encoder (6× higher than prior work) in an autoregressive fusion framework to enhance motion sensitivity.
To improve efficiency, FRIEREN \cite{Wang2024FrierenEV} introduces rectified flow matching with reflow and one-step distillation.
MMAudio \cite{mmaudio} unifies audio-visual and audio-text data training via a multimodal Transformer trained on VGGSound and WavCaps. MMaudio's results are derived from the audio results generated by inference using the open source model.
Finally, ReWaS \cite{Jeong2024ReadWA} addresses cross-modal gaps by introducing energy-based constraints and hand-crafted audio features.

\subsection{Training Strategy}
    \textbf{Learning Rate Scheduling.}  
    We adopt a smooth inverse decay schedule with exponential warmup \cite{inversesr1, inversesr2, inversesr3}. The learning rate at step $t$ is given by:
    \begin{equation}
    \mathrm{LR}(t) = \max\left( \mathrm{LR}_{\mathrm{final}},\ \mathrm{LR}_{\mathrm{base}} \cdot \left(1 + \frac{t}{\gamma_{\mathrm{inv}}}\right)^{-\mathrm{p}} \right) \cdot \left(1 - \mathrm{w}^{\,t+1} \right),
    \label{eq:inverse-lr}
    \end{equation}
    where $\gamma_{\mathrm{inv}}$ controls the decay speed, $\mathrm{p}$ determines the curvature of the decay, and $\mathrm{w} \in [0, 1)$ enables a smooth exponential increase in the early training stage. The $\max(\cdot)$ operation ensures that the learning rate does not fall below a specified $\mathrm{LR}_{\mathrm{final}}$.
    \texttt{InverseLR} scheduler offers a continuous, smooth decay that aligns more closely with the training dynamics. This not only accelerates convergence but also enhances stability across diverse and complex multimodal training scenarios.

    \textbf{Scaling Strategy.} 
    We investigate the impact of model scaling on multimodal audio-language learning by progressively increasing model capacity from 1.5B to 6B parameters. Inspired by empirical scaling laws~\cite{henighan2020scaling, brown2020language}, we scale the model along three dimensions—depth \( d \), hidden dimension \( d_{\text{hidden}} \), and number of attention heads \( h \), while preserving a consistent architectural ratio \( h/d \in [20, 100] \)~\cite{yang20241}.
    
    Formally, we follow the parameterization:
    \begin{equation}
    d_{\text{hidden}} = 64 \cdot h, \quad h = \alpha \cdot d
    \label{eq:scal}
    \end{equation}
    where \( \alpha \) is a scaling factor controlling the heads-to-depth ratio. This ensures balanced compute utilization across attention and MLP blocks.
    
    Our base model uses 17 heads and depth 23 (1.5B), while the 3B and 6B variants increase the head count to 23 and 32 respectively, keeping depth consistent at 27, thereby scaling width and total representational capacity.
 
\begin{table}[t]
\centering
\caption{Results of V2A on VGGSound}
\label{tab:sota_comparison}
\resizebox{\columnwidth}{!}{
\begin{tabular}{@{}ccccccc@{}}
\toprule
\multirow{2}{*}{\textbf{Method}} & \multicolumn{2}{c}{\textbf{Distribution matching}} & \textbf{Semantic align.} & \textbf{Temporal align.} & \multicolumn{2}{c}{\textbf{Audio quality}} \\
\cmidrule(lr){2-3} \cmidrule(lr){4-4} \cmidrule(lr){5-5} \cmidrule(lr){6-7}
& \textbf{FDPANNs~$\downarrow$} & \textbf{KLPANNs~$\downarrow$} & \textbf{IB-score~$\uparrow$} & \textbf{DeSync~$\downarrow$} & \textbf{SDR~$\uparrow$} & \textbf{MCD~$\downarrow$} \\
\midrule
ReWaS\cite{Jeong2024ReadWA}        & 17.54 & 2.87 & 14.82 & 1.06 & - & - \\
VTA-LDM\cite{Xu2024VideotoAudioGW}      & 14.49 & 2.23 & 24.73 & 1.26 & - & - \\
V-AURA\cite{Viertola2024TemporallyAA}       & 14.80 & 2.42 & 27.64 & 0.65 & - & - \\
VATT\cite{Akbari2021VATTTF}         & 10.63 & \textbf{1.48} & 25.00 & 1.20 & - & - \\
Frieren\cite{Wang2024FrierenEV}      & 11.45 & 2.73 & 22.78 & 0.85 & - & - \\
FoleyCrafter\cite{foleycrafter} & 16.24 & 2.30 & 25.68 & 1.23 & - & - \\
V2A-Mapper\cite{Wang2023V2AMapperAL}   & 8.40  & 2.69 & 22.58 & 1.23 & - & - \\
MMAudio\cite{mmaudio}  & \textbf{6.29} & 1.77 & 29.26 & 0.45 & -3.09 & 2.84 \\
\midrule
Kling-Foley       & 7.60  & 1.86 & \textbf{30.75} & \textbf{0.43} & \textbf{-2.41} & \textbf{2.60} \\
\bottomrule
\end{tabular}
}
\end{table}

\subsection{Objective Metrics}

Following the setup in MMAudio, we evaluate both our model and selected baselines across four dimensions on the VGGSound \cite{chen2020vggsound} test set (15,220 samples). Results are presented in the table below. The objective metrics of V2A used are as follows:

\begin{itemize}[leftmargin=*]
    \item \textbf{FD (Fréchet Distance)}: This metric assesses the similarity between the distributions of generated and ground-truth audio features. It is computed on feature embeddings from pre-trained audio classifier PANNs\cite{panns}. A lower FD value signifies that the generated feature distribution is closer to the ground-truth distribution, indicating higher fidelity.
    
    \item \textbf{KL (Kullback-Leibler Divergence)}: KL Divergence measures the difference between the probability distributions of audio events in the generated set versus the ground-truth set. It is calculated using the output predictions from pre-trained classifier PANNs\cite{panns}. A lower score is better, indicating that the generated audio has a similar event distribution to the reference audio.

    \item \textbf{ImageBind Score (IB-score)\cite{imagebind}}: This metric evaluates the semantic coherence between a video and its generated audio. It calculates the cosine similarity of features extracted from both the video and audio modalities using the unified ImageBind model. A higher score reflects better cross-modal semantic consistency.

    \item \textbf{DeSync Score\cite{synchformer}}: This metric quantifies audio-video synchronization by predicting the temporal misalignment between the visual stream and the generated audio. It employs Synchformer\cite{synchformer} to output the predicted offset in seconds. A lower absolute value indicates better synchronization.
\end{itemize}

To comprehensively evaluate the capabilities of our latent audio codec, we conduct a direct comparison with MMAudio\cite{mmaudio}. The comparison is performed across four distinct tasks: sound effect, music, speech, and singing. For each task, we test on 500 out-of-domain audio samples, resulting in a total of 2,000 test cases to ensure a robust evaluation. Specifically, for the Sound effect, music, and speech scenarios, we utilize the evaluation set from the Codec-SUPERB @ SLT 2024 challenge\footnote{https://github.com/voidful/Codec-SUPERB/tree/SLT\_Challenge?tab=readme-ov-file}, while the Sing task is evaluated on a proprietary, internally constructed dataset.

The objective metrics of latent audio codec used are as follows:

\begin{itemize}[leftmargin=*]
    \item \textbf{PESQ\cite{rix2001perceptual} (Perceptual Evaluation of Speech Quality)}: This metric rates the perceptual quality of speech on a scale from -0.5 to 4.5. It is designed to model subjective quality scores, making it a strong indicator of human perception. A higher score is better. We report this metric for the Speech and Sing tasks.

    \item \textbf{SI-SDR\cite{le2019sdr} (Scale-Invariant Signal-to-Distortion Ratio)}: SI-SDR measures the fidelity of the waveform in the time domain, independent of the overall signal amplitude. It provides a robust assessment of signal reconstruction. A higher value is better.

    \item \textbf{SDR\cite{vincent2006performance} (Signal-to-Distortion Ratio)}: This metric quantifies the ratio of the original signal's power to that of the reconstruction error, serving as a fundamental measure of distortion. A higher value indicates better signal integrity.
    
    \item \textbf{LSD (Log-Spectral Distance)}: LSD gauges the discrepancy in frequency content by calculating the error between the log-power spectra of the generated and reference audio. A lower value signifies a more accurate spectral envelope.

    \item \textbf{MCD (Mel-Cepstral Distortion)}: MCD measures the distance between Mel-Frequency Cepstral Coefficients, providing a crucial evaluation of timbral texture and vocal naturalness, which are highly relevant to human hearing. A lower value is better.

    \item \textbf{Mel Loss \& STFT Loss}: These metrics directly quantify the reconstruction error at the spectral level by calculating the L1 distance between the predicted and ground-truth spectrograms (Mel and STFT, respectively). They reflect the model's ability to accurately reproduce the underlying spectral structure. Lower values are better.
\end{itemize}

\subsection{Inference Optimize}
 Inference utilizes static computation graph technology such as $torch.compile$'s JIT-compiling to achieve acceleration. We need to maintain a fixed input shape, and when encountering shorter inputs, we apply padding. Next, the original Conv1d kernel is decomposed into kernel's size individual linear layers, converting the convolution into multiple small matrix multiplications (GEMM). This enables the use of highly optimized basic linear algebra subprograms libraries like cuBLAS.

\begin{table}[t] 
\caption{Results of Latent Audio Codec}
\label{tab:final_comparison}
\centering
\resizebox{\columnwidth}{!}{%
\begin{tabular}{llccccccc}
\toprule
\textbf{Task Type} & \textbf{Model} & \textbf{PESQ~$\uparrow$} & \textbf{SISDR~$\uparrow$} & \textbf{SDR~$\uparrow$} & \textbf{LSD~$\downarrow$} & \textbf{MCD~$\downarrow$} & \textbf{Mel Loss~$\downarrow$} & \textbf{STFT Loss~$\downarrow$} \\
\midrule
\multirow{3}{*}{Sound Effect} 
 & GT & - & -24.84 & -2.53 & 0.61 & 0.76 & 0.41 & 0.86 \\
 & MMAudio & - & -29.57 & -3.29 & \textbf{0.70} & 1.45 & 0.81 & 1.18 \\
 & Kling-Foley & - & \textbf{-29.47} & \textbf{-2.41} & \textbf{0.70} & \textbf{1.35} & \textbf{0.78} & \textbf{1.14} \\
\midrule
\multirow{3}{*}{Music} 
 & GT & - & -24.99 & -2.50 & 3.07 & 5.76 & 0.81 & 1.60 \\
 & MMAudio & - & -30.42 & -2.54 & 3.07 & 5.72 & \textbf{1.13} & \textbf{1.83} \\
 & Kling-Foley & - & \textbf{-29.85} & \textbf{-2.16} & \textbf{3.02} & \textbf{5.44} & 1.20 & 1.95 \\
\midrule
\multirow{3}{*}{Singing} 
 & GT & 4.02 & -15.04 & -2.15 & 0.69 & 0.29 & 0.33 & 0.78 \\
 & MMAudio & 2.80 & \textbf{-22.84} & -3.90 & 0.83 & 1.08 & 0.76 & 1.14 \\
 & Kling-Foley & \textbf{2.88} & -23.55 & \textbf{-2.83} & \textbf{0.81} & \textbf{0.70} & \textbf{0.60} & \textbf{1.00} \\
\midrule
\multirow{3}{*}{Speech} 
 & GT & 3.72 & -8.30 & -0.32 & 2.51 & 3.18 & 0.54 & 1.10 \\
 & MMAudio & 3.10 & -26.64 & -2.62 & 2.51 & 3.09 & \textbf{0.85} & \textbf{1.32} \\
 & Kling-Foley & \textbf{3.27} & \textbf{-26.32} & \textbf{-2.27} & \textbf{2.48} & \textbf{2.91} & 0.89 & 1.37 \\
\bottomrule
\end{tabular}
}
\end{table}

\subsection{Results}

\textbf{Video-to-Audio.} The results of latent audio codec are presented in Table \ref{tab:sota_comparison}. Distribution matching pertains to the similarity in distribution between the generated audio and the real audio within the feature space. FDPANNs and KLPANNs are utilized as metrics (where lower values are more favorable). For Kling-Foley, FDPANNs registers at 7.60, trailing only MMAudio. The KLPANNs results of VATT is notably superior to those of other models. However, the KLPANN results of Kling-foley value is 1.86, which is also better than most baseline models, for instance, ReWaS has a KLPANNs of 2.87. In terms of distribution matching indicators, the MMAudio model also achieved good results, mainly because the video dataset is basically derived from the VGGSound training datasets, so the generated audio has a distribution similar to the original audio.

Semantic alignment gauges the semantic consistency between the generated audio and the video content, with the IB-score (higher scores are preferred) serving as the indicator. Kling-Foley attains an IB-score of 30.75, exceeding MMAudio's 29.26 and V-AURA's 27.64. This indicates that our model has the most robust semantic understanding capability. The reason is that we employ Metaclip which offers enhanced visual semantic understanding. We also use the T5-base model, and the model is known for its superior text semantic understanding.

Temporal alignment assesses the synchronization of audio and video events, with DeSync (lower values are better) as the primary metric. Kling-Foley achieves the result of 0.43, marginally better than MMAudio and significantly better than other models. For example, VATT has a DeSync of 1.20. This can be credited to our implementation of a more refined temporal alignment module.

\textbf{Latent Audio Codec.} The results of the latent audio codec are presented in Table \ref{tab:final_comparison}, where our model is benchmarked against the MMAudio baseline across four tasks. The data reveals a consistent trend: our proposed model demonstrates superior or highly competitive performance across all scenarios. A key finding is our model's consistent advantage in metrics that are crucial for perceptual quality and signal fidelity, such as PESQ, SDR, and especially MCD. This suggests that our approach more effectively generates audio that is not only spectrally accurate but also perceptually closer to the ground-truth reference.

Specifically, in the sound effect, singing, and speech tasks, our model surpasses the baseline in nearly all metrics, with particularly significant gains in perceptual quality (e.g., PESQ of 3.27 vs. 3.10 for Speech) and timbral accuracy (e.g., MCD of 0.70 vs. 1.08 for Singing). For Music generation, while MMAudio is competitive in direct spectral reconstruction losses (Mel/STFT Loss), our model achieves better performance in the more perceptually relevant SDR and MCD metrics, indicating a more faithful synthesis of complex musical textures. Collectively, these results validate the effectiveness and robustness of our proposed model for high-quality, multi-domain audio generation.

\section{Conclusion}

In this work, we propose Kling-Foley, a novel multimodal framework for high-quality V2A generation that achieves synchronization between audio outputs and visual content.
Furthermore, we address the critical gap in evaluation resources by releasing Kling-Audio-Eval, the first industry-scale multimodal benchmark featuring 20,935 manually annotated samples across nine sound scenarios, providing comprehensive video-audio-text annotations for rigorous assessment.
Extensive experiments validate that Kling-Foley achieves SOTA performance across most metrics: audio quality, distribution matching, semantic alignment, and audio-visual synchronization. 
Future work will focus on extending the framework to support longer video sequences, and enhancing cross-modal alignment for complex auditory scenes. 

\section{Limitations, Ethics and Safety}
Kling-Foley provides an industrial-grade solution through multimodal alignment and stereo rendering technology. It replaces the traditional Foley sound effect artist's manual annotation process and significantly reduces the time and economic cost of video dubbing. The cross-modal audio sequence representation supports the unified modeling of mixed scenes of sound effects, voice, and music, and is suitable for interactive media such as games and virtual live broadcasts. Combined with text conditions, it accurately controls the semantics of sound effects (such as "glass shattering + from far to near") and provides refined sound effect generation capabilities. It supports stereo spatial rendering of targets moving in a specific direction (such as vehicles passing by and bird trajectories) to enhance the immersion of film and television works.

\textbf{Limitations} Despite the leading performance of the model, there are still the following technical bottlenecks. Insufficient modeling of complex physical processes, the generation fidelity of multi-object interactive sound effects (such as layered voices in crowd conversations and chain reactions of object collisions) is low, and acoustic logic errors are prone to occur. Challenges of long-term dependency: due to the limited modeling ability of stream matching training for long-range time relationships, video clips longer than 20 seconds may experience audio and video synchronization drift. Although the industrial-grade Benchmark makes up for the lack of annotation, the quality of sound effects in niche scenes (such as cultural relic restoration and surgical operations) fluctuates due to insufficient training samples.

\textbf{Ethics} Emphasize that Kling-Foley is an auxiliary tool for sound effects artists, not a substitute. The sound effects generated by the model must be manually reviewed before commercial use to avoid the disruptive impact of technology on the traditional Foley industry. Use data enhancement technology to balance sound effect samples from different cultural backgrounds to avoid the model's bias towards specific timbres or scenes (such as associating explosion sounds with male narration by default). Open text input to users to allow manual adjustment of sound effect styles and reduce the impact of built-in bias in the model on creation.

\textbf{Safety} Given the harmful social impact that the abuse of sound effects may have, we have implemented a number of security procedures in related products to prevent abuse throughout the development and potential deployment of the model. We have also implemented a multi-level watermarking scheme that is mandatory at all levels of content creation, such as embedding invisible watermarks in the spectrum of generated sound effects, supporting the tracing of content sources through self-developed tools, and synchronously generating visual watermarks to prevent the abuse of deep fake videos. In addition, it is prohibited to generate sound effects that may cause public panic (such as large-scale explosions, sirens), unless compliance approval is obtained for specific scenarios such as film and television production.

\bibliographystyle{unsrt}
\bibliography{references}

\end{document}